\documentclass[showpacs,notitlepage,floatfix,citeautoscript,superscriptaddress,prb,reprint]{revtex4-2}
\usepackage{graphicx}
\usepackage{braket}
\usepackage{color,soul}

\usepackage{mathrsfs}
\usepackage{times}

\usepackage[dvipsnames,svgnames,table]{xcolor}
\usepackage{amsmath}
\usepackage{bm}
\usepackage[T1,OT1]{fontenc}

\usepackage{hyperref}
\usepackage{fancyhdr}
\usepackage{float}

\hypersetup{
pdfstartview={FitH},
colorlinks=true,    
linkcolor=NavyBlue, 
citecolor=Maroon,   
filecolor=NavyBlue, 
urlcolor=NavyBlue   
}

\begin{document}

\title{The robustness of spin-polarized edge states in a two-dimensional
topological semimetal without inversion symmetry}

\author{Jos\'e.~D. Mella}
\email{jmellariquelme@gmail.com}
\affiliation{Departamento de F\'isica, Facultad de Ciencias F\'isicas y Matem\'aticas,
  Universidad de Chile, Santiago, Chile}

\author{Luis E. F. Foa Torres}
\affiliation{Departamento de F\'isica, Facultad de Ciencias F\'isicas y Matem\'aticas,
  Universidad de Chile, Santiago, Chile}

\begin{abstract}

Three-dimensional topological gapless phases have attracted significant attention due to their unique electronic properties. A flagship example are Weyl semimetals, which require breaking time-reversal or inversion symmetry. In two dimensions, the dimensionality reduction requires imposing an additional symmetry, thereby weakening the phase. Like its three-dimensional counterpart, these ``\textit{two-dimensional Weyl semimetals}'' present edge states directly related to Weyl nodes. The direct comparison with the edge states in zigzag-like terminated graphene ribbons is unavoidable, offering the question of how robust these states are and their differences. Here we benchmark the robustness of the edge states in two-dimensional Weyl semimetals without inversion symmetry with those present in zigzag graphene ribbons. Our results show that, despite having a similar electronic bandstructure, the edge states of two-dimensional Weyl semimetal are more robust against vacancies than graphene ribbons. We attribute this enhanced robustness to a crucial role of the spin degree of freedom in the former case.

\end{abstract}

\maketitle
\section{Introduction}
\label{sec:intro}

From the first theoretical proposal of a topological insulator~\cite{kane2005z,kane2005quantum,bernevig_quantum_2006} and thrilling experiments that followed~\cite{konig_quantum_2007,hsieh_topological_2008}, different classes of topological phases have been studied~\cite{hasan2010colloquium,qi2011topological,yan2012topological,munoz2018topological,groning2018engineering,liu2021spin}. Topological but gapless, Weyl and Dirac semimetals (WSM and DSM) have recently attracted significant attention due to their unique electronic properties. These include ultrahigh mobility~\cite{shekhar2015extremely,liang2015ultrahigh}, negative magnetoresistance\cite{huang2015observation,wang2016gate,arnold2016negative}, colossal photo-voltaic response~\cite{osterhoudt2019colossal,ma2019nonlinear}, and enhanced catalytic behavior~\cite{rajamathi2017weyl}. Their electronic bandstructure consists of the conduction and valence bands cross linearly at isolated fourfold (twofold) degenerate points called Dirac (Weyl) nodes in the Brillouin zone. The premier tool for experimentally identifying bulk and surface band structure is angle-resolved photoemission spectroscopy (ARPES)~\cite{lv2015observation,lv2015experimental,lv2015observation2,xu2015discovery,xu2015discovery2}. 

Co-dimensional analysis in three-dimensional (3D) systems shows a DSM phase between the topological and trivial insulator phase when time-reversal ($\mathcal{T}$) and inversion ($\mathcal{I}$) symmetry are present~\cite{murakami2007phase}. When some perturbation breaks one of these symmetries, the Dirac node splits into two Weyl nodes, obtaining a $\mathcal{I}$~\cite{huang2015weyl,weng2015weyl,xu2015discovery,lv2015experimental} or $\mathcal{T}$~\cite{wan2011topological,xu2011chern,xu2018topological} symmetry breaking WSM phase. A no-go theorem~\cite{nielsen1983adler} forces Weyl nodes to appear in pairs with opposite chirality~\cite{yu2016determining,ma2017direct} with consequences in the band structure: The appearance of surface states called Fermi arcs. Their topological nature can be traced to the non-zero Chern number in two-dimensional planes between two Weyl nodes with opposite chirality~\cite{wan2011topological,xu2011chern,balents2011weyl}.

\begin{figure}[t!]
    \centering
    \includegraphics[width=0.7\columnwidth]{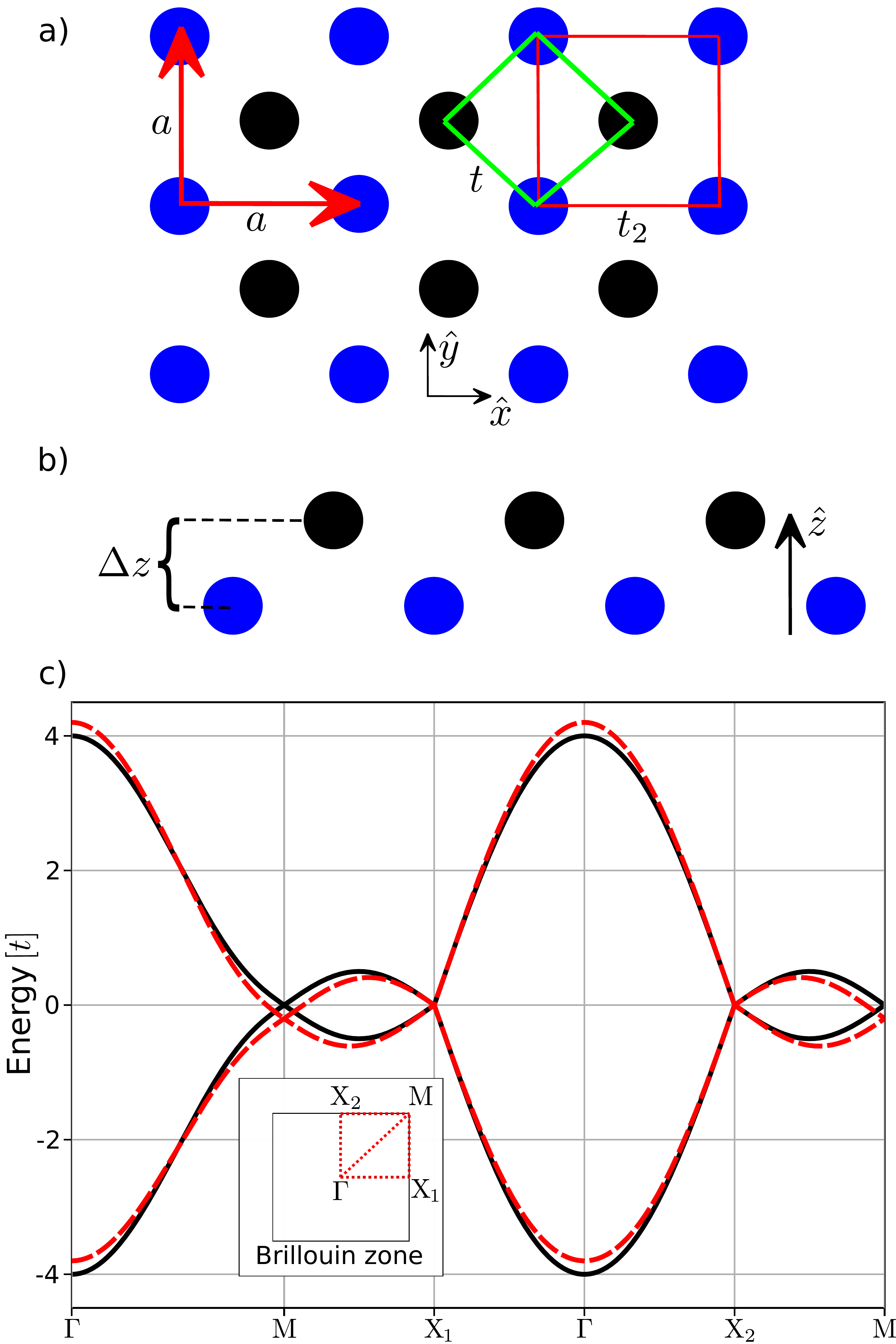}
    \caption{a) Top view of the system where red square is the unit cell and green/red thick line shows the first/second nearest-neighbor interaction. b) The lateral view, where the two equal but nonequivalent sites have a separation perpendicular to the plane. c) Electronic band structure along high symmetry points of the Young and Kane model\cite{young2015dirac} with $t=2$, $t_2=0.03t$ and $t^{\text{SO}}=0.25t$. Thick/dotted line is without/with second nearest neighbors interaction, where the dotted line shows the particle-hole symmetry breaking effects. }
    \label{fig:setup}
\end{figure}

Due to the dimensionality reduction~\cite{murakami2007phase}, and in contrast to their three-dimensional counterparts, a two-dimensional metallic phase requires imposing an additional symmetry. In what follows, we call such phases two-dimensional (2D) WSM phase, or DSM phase, depending on the degeneracy of the nodes. These phases can be grouped in two categories: \textit{i}) When nodes occur by an accidental band crossing along the invariant extra symmetry line and can be pairwise annihilated without breaking the additional symmetry~\cite{baik2015emergence,zhang2014prediction,park2017classification,ahn2017unconventional}. Surface-dopped black phosphorus is a good example, where two Dirac nodes are predicted and confirmed experimentally~\cite{kim2017two}.  \textit{ii}) When nodes occur by a crystal symmetry and can not be pairwise annihilated~\cite{young2015dirac,jin2020two}, such as the model proposed by Young and Kane~\cite{young2015dirac}, the non-symmorphic symmetry $\{g|\textbf{t}\}$, formed by a point group symmetry plus a fractional translation of the Bravais lattice vector \textbf{t}, ensures the appearance of a Dirac node. Because of this additional symmetry requirement, one may wonder how robust are the ensuing edge states. Indeed, these states are reminiscent of the edge states present in zigzag graphene ribbons\cite{matveeva2019edge,luo2020two}, which despite having an origin that could be traced to a Zak phase, turn out to be very fragile to disorder and imperfections. A natural question thus concerns the robustness of the edge states in 2D WSMs.

Here we address how robust are the edge states in 2D WSM and compare them with another paradigmatic case: the edge states of zigzag ribbons. We illustrate our conclusions by using the model proposed by Young and Kane~\cite{young2015dirac}, where we break symmetries until we isolate four Weyl nodes without $\mathcal{I}$ symmetry. Specifically, we compute the band structure, including 1D Fermi arcs, and study their origin. Using scattering theory, we also calculate the transport properties in a two-terminal setup and examine the robustness against vacancies. Scrutinizing the different contributions to the transmission and reflection probabilities allows us to interpret the transport results that we benchmark against graphene with a zigzag termination. Our analysis reveals a subtle effect of symmetries and spin that render the edge states in 2D WSMs more robust than those of zigzag-terminated graphene.

\section{Two Dimensional Dirac semi-metal model}

Our starting point is the two-dimensional DSM model introduced by Young \textit{et al.}~\cite{young2015dirac} consists of a square lattice (P4/$nmm$ group) with two $s$ orbitals in the unit cell and a separation between them in the $\hat{z}$-direction (fig.~\ref{fig:setup}a and b). The effective Hamiltonian, including spin-orbit coupling and the second nearest-neighbor hoppings, is

\begin{eqnarray}\label{eq:Hamiltonian}
H(\textbf{k})&=&2t\tau_x\cos\left(\frac{k_x}{2}\right)\cos\left(\frac{k_y}{2}\right)+ \nonumber\\
&+&t_2(\cos(k_x)+\cos(k_y))+ \nonumber \\
&+&t^{\text{SO}}\tau_z(\sigma_y\sin(k_x)-\sigma_x\sin(k_y)),
\end{eqnarray}

with $\tau(\sigma)$ Pauli matrices for the orbital (spin) degree of freedom. The system preserves $\mathcal{T}$ and  $\mathcal{I}$ symmetry, together with nonsymmorphic symmetries screw axis $\{C_{2\hat{x}}|\frac{1}{2}0\}$ and $\{C_{2\hat{y}}|0\frac{1}{2}\}$ and glide mirror plane $\{M_{\hat{z}}|\frac{1}{2}\frac{1}{2}\}$. 

The nonsymmorphic symmetry $\{g|\textbf{t}\}$ plays an important role in describing the system's electronic properties. Without other symmetries, a pair of bands must intersect an odd number of times along the invariant line (or plane) in the Brillouin zone, satisfying $g\textbf{k}=\textbf{k}$. The presence of time-reversal ($\mathcal{T}=i\sigma_y\mathcal{K}$ with $\mathcal{T}^2=-1$) and inversion symmetries ($\mathcal{I}=\tau_x$) implies that the crossing points due to the nonsymmorphic symmetry must be at time-reversal invariant momenta (TRIM)~\cite{young2015dirac}. In this case, the invariant line/plane corresponds to $\{C_{2\hat{x}}|\frac{1}{2}0\}$, $\{C_{2\hat{y}}|0\frac{1}{2}\}$ and  $\{M_{\hat{z}}|\frac{1}{2}\frac{1}{2}\}$ symmetry is $k_y=0,\pm \pi$, $k_x=0,\pm \pi$ and the two dimensional Brillouin zone plane ($k_z=0$). This yields three intersection points in $X_1=(\pi,0)$, $X_2=(0,\pi)$ and $M=(\pi,\pi)$, where the Dirac nodes are arisene~\cite{matveeva2019edge,young2015dirac}, as it can see with a thick line in the electronic bandstructure in Fig.~\ref{fig:setup}c. $t_2 \neq 0$ breaks particle-hole symmetry, changing Dirac nodes energies of $M$ relative to the $X_1,X_2$ point (dotted line in fig.~\ref{fig:setup}c), essential for the transport properties that we explain in the transport properties section. The symmetry protection of Dirac nodes is given by $\mathcal{T}$ and $\mathcal{I}$ symmetry, together with $\{M_{\hat{z}}|\frac{1}{2}\frac{1}{2}\}$ for $X_1,X_2$ node, $\{C_{2\hat{x}}|\frac{1}{2}0\}$ for $X_1,M$ node and $\{C_{2\hat{y}}|0\frac{1}{2}\}$ for $X_2,M$ node.

\begin{figure}[ht!]
    \centering
    \includegraphics[width=0.8\columnwidth]{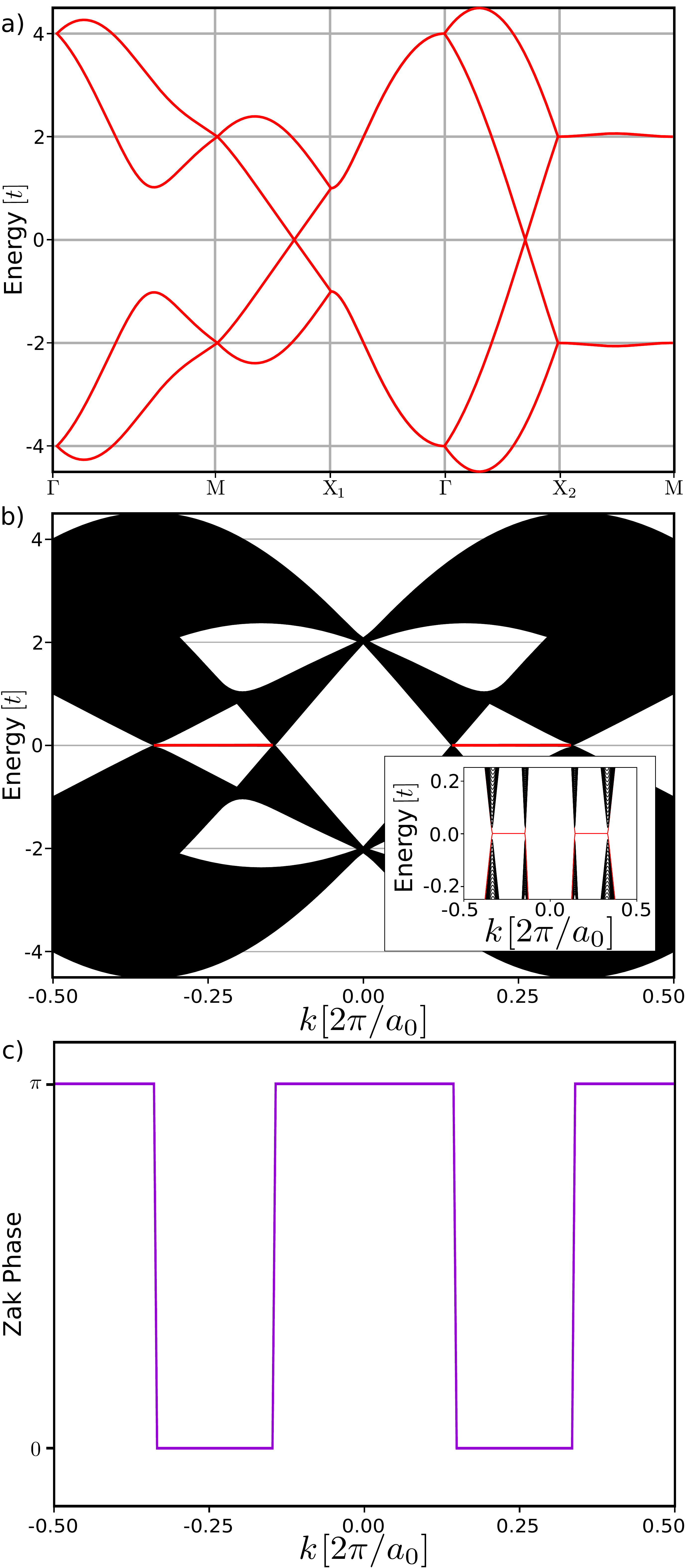}
    \caption{a) Band structure of $H'$, where the Weyl nodes are near $M$ and $X_2$. b) Band structure of a ribbon $H'(k_y)$ with 300 sites in $\hat{x}$-direction, where the red lines are the one dimensional (1D) Fermi arcs, which can clearly show in the inset panel in a zoom near Fermi level. c) The Zak phase in 1D lines in $k_x$ for any constant $k_y$ value in Brillouin Zone, where a difference of $\pi$ value shows a change in the edge states (from here, $k$ refer to $k_y$). The parameter are the same of fig.~\ref{fig:setup} together with $v_1=0.5t,v_2=-t,v_3=t$ and $t_2=0$}.
    \label{fig:edge}
\end{figure}

\section{2D Weyl semimetal Model}
By adding new terms to the Hamiltonian of the previous Section, Eq.~\ref{eq:Hamiltonian}, one gets a two-dimensional Weyl semimetal. In the following, we explain the new Hamiltonian terms that need to be introduced and their consequences in the electronic structure. In order to reduce the number of Dirac points near the Fermi energy, the first Hamiltonian term to be added is 

\begin{eqnarray}
H_1(\textbf{k})=v_1\sin\left(\frac{k_x}{2}\right)\cos\left(\frac{k_y}{2}\right)\tau_y,
\end{eqnarray}

whose consequence in the lattice model is a strain that changes the angle between the two basis vectors. While $\{C_{2\hat{y}}|0\frac{1}{2}\}$ symmetry is preserved, $\{C_{2\hat{x}}|\frac{1}{2}0\}$ and $\{M_{\hat{z}}|\frac{1}{2}\frac{1}{2}\}$ are broken, destroying the protection of Dirac node in $X_1=(\pi,0)$, opening a band-gap and leaving the Dirac nodes in $X_2$ and $M$ without perturbation. The next term is a spin-orbit interaction term, 

\begin{eqnarray}
H_2(\textbf{k})&=&v_2\cos\left(\frac{k_x}{2}\right)\sin\left(\frac{k_y}{2}\right)\tau_x\sigma_y,
\label{eq:rashba1}
\end{eqnarray}
where $\{M_{\hat{z}}|\frac{1}{2}\frac{1}{2}\}$ is broken together $\mathcal{I}$ symmetry and can arise because of different dipole moment configurations at each site\cite{young2015dirac}. The Dirac nodes in $X_2$ splits into two Weyl nodes on the symmetry invariant line $k_x=0$ related to $\{C_{2\hat{y}}|0\frac{1}{2}\}$. At $M$, $H_2\rightarrow 0$, leaving its Dirac node undisturbed against $H_2$. Usually, the symmetry constrains forces the system to have two Dirac nodes, being impossible to isolate only one ~\cite{young2015dirac,matveeva2019edge}. In consequence and in the spirit of reducing the number of bands near the Fermi level, we introduce a new term of an analogous origin of eq.~\eqref{eq:rashba1} that converts the Dirac node in $M$ into two Weyl nodes along $k_x=\pi$, associated to $\{C_{2\hat{y}}|0\frac{1}{2}\}$,

\begin{eqnarray}
H_3(\textbf{k})&=&v_3\sin\left(\frac{k_x}{2}\right)\sin\left(\frac{k_y}{2}\right)\tau_y\sigma_y.
\end{eqnarray}

The 2D Weyl semimetal model consists of the sum of all the previously mentioned terms:

\begin{equation}
\label{eq:weyl-hamiltonian}
H'(\textbf{k})=H(\textbf{k})+H_1(\textbf{k})+H_2(\textbf{k})+H_3(\textbf{k}), 
\end{equation}

which has $\{C_{2\hat{y}}|0\frac{1}{2}\}$ and $\mathcal{T}$ symmetries. The resulting electronic band structure with particle-hole symmetry can be seen in fig.~\ref{fig:edge}a. There are two Weyl nodes in $k_x=0,\pi$, consistent with the crossing band due to the nonsymmorphic symmetry mentioned in the previous section. The protection of the 2D Weyl nodes is related to the symmetries that generated the crossing bands.

 \begin{figure}[ht!]
    \centering
    \includegraphics[width=0.9\columnwidth]{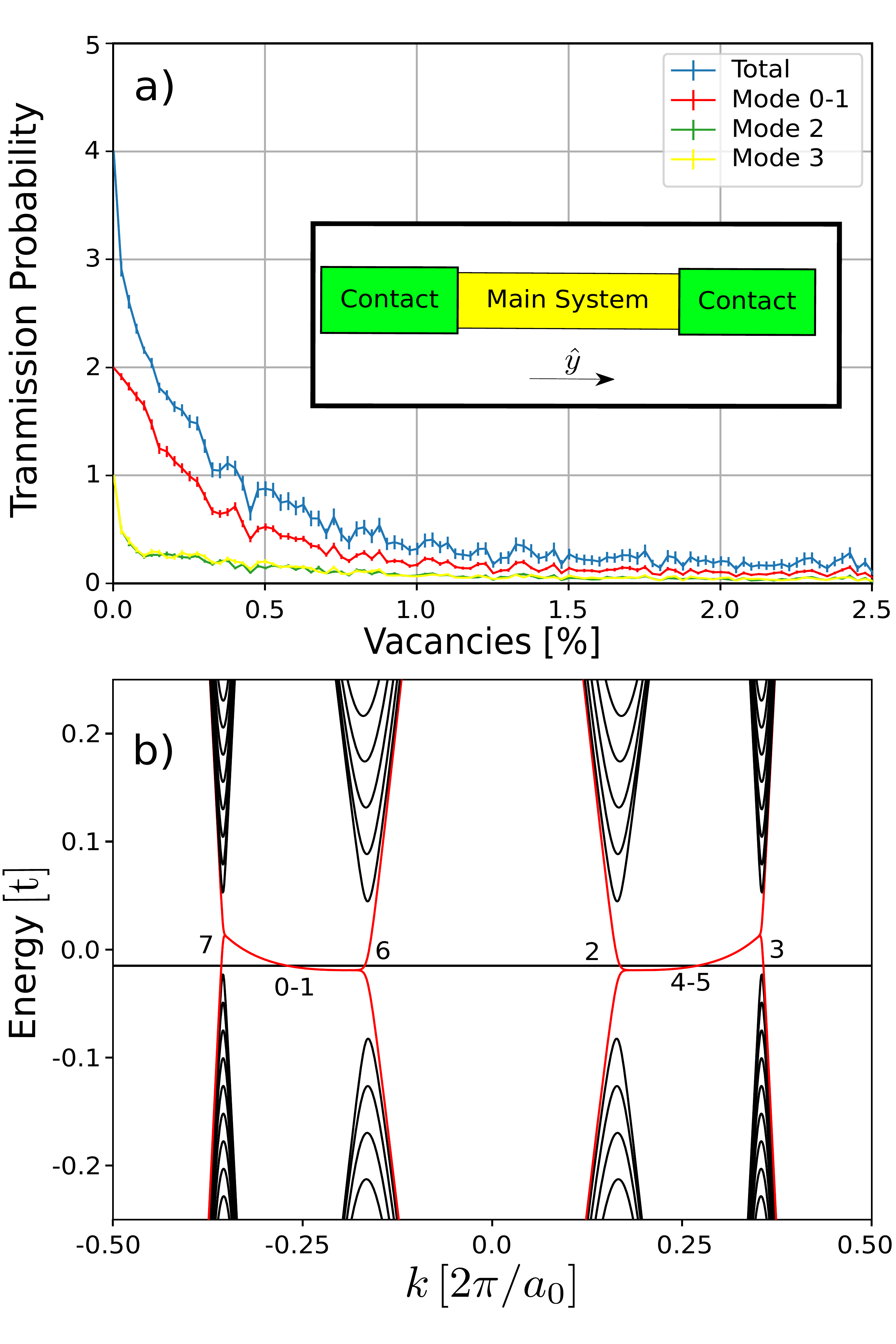}
    \caption{a) Transmission results for every mode showed in b) involved in the transport calculation of the two-contact 2D WSM system.  A $100\times 1000$ main system is considered in a scheme shown in the subset with its corresponding contact (with the same interaction of the main system). Four bands are involved in the transmission process for an incoming wave from left to right contact (referring to the subset's scheme). Two double degenerate edge states channel (0-1) and two bulk-like channels (2 and 3).}
    \label{fig:contribution}        
\end{figure}

\section{Edge states}

The surface states of WSMs, the Fermi arcs, bridge two Weyl nodes with opposite monopole charges in the Brillouin zone, as long as the projection of both nodes is not at the same point. In 3D, the Fermi arcs result from the non-zero Chern number calculated along planes in the Brillouin zone perpendicular to the lines connecting Weyl nodes with opposite chirality. In their two-dimensional counterpart, a quantized value of the Zak Phase suggests where Fermi arcs appear. The Zak phase is calculated along a line in the Brillouin zone and contrasts with the Chern numbers calculated on planes of a three-dimensional Brillouin zone. The band structure of a ribbon, with periodical boundary condition in $k_y$ for our model, is shown in Fig.~\ref{fig:edge}b), where four Weyl nodes projections arise. Near Fermi energy, both flat bands bridge one Weyl node near M with another one near X$_2$, bearing opposite topological charges, which we call one dimensional (1D) Fermi arcs. To confirm the appearance of the 1D Fermi arc, for a constant value of $k_y$, we calculated the Zak phase through $k_x$ correspond to the first Brillouin zone (fig.~\ref{fig:edge}c), where the appearance of the 1D Fermi arcs depends on a change in the $\pi$ value of Zak Phase. The presence of particle-hole symmetry constrains the energy of Weyl nodes to be the same, obtaining two flat bands that bridge those nodes. The second nearest neighbor interaction ($t_2$) breaks that symmetry, changing the energy of the Weyl nodes, giving a little dispersion to the edge states that depend on the strength of the interaction. Examples of particle-hole symmetry break can be seen in fig.~\ref{fig:contribution}b, which will be helpful for the transport calculation in the next section.

At this point, the comparison between 1D Fermi arcs and the flat bands of zigzag terminated graphene is natural~\cite{deSousa2021}. In graphene, the Dirac cones in $K$ and $K'$ have opposite chirality related to the pseudo-spin~\cite{torres2013introduction}. $\mathcal{T}$ and $\mathcal{I}$ are preserved, making all the bands double degenerate. The flat band has a topological origin in the zigzag edge because the winding number is different from zero. However, the robustness of these states is limited by that of the normal mode decomposition and the chiral symmetry~\cite{delplace2011zak}, which turns out to be very weak against onsite disorder, spin-orbit coupling, or some perturbation that break $\mathcal{T}$ or $\mathcal{I}$.

\begin{figure*}[t!]
    \centering
    \includegraphics[width=0.8\linewidth]{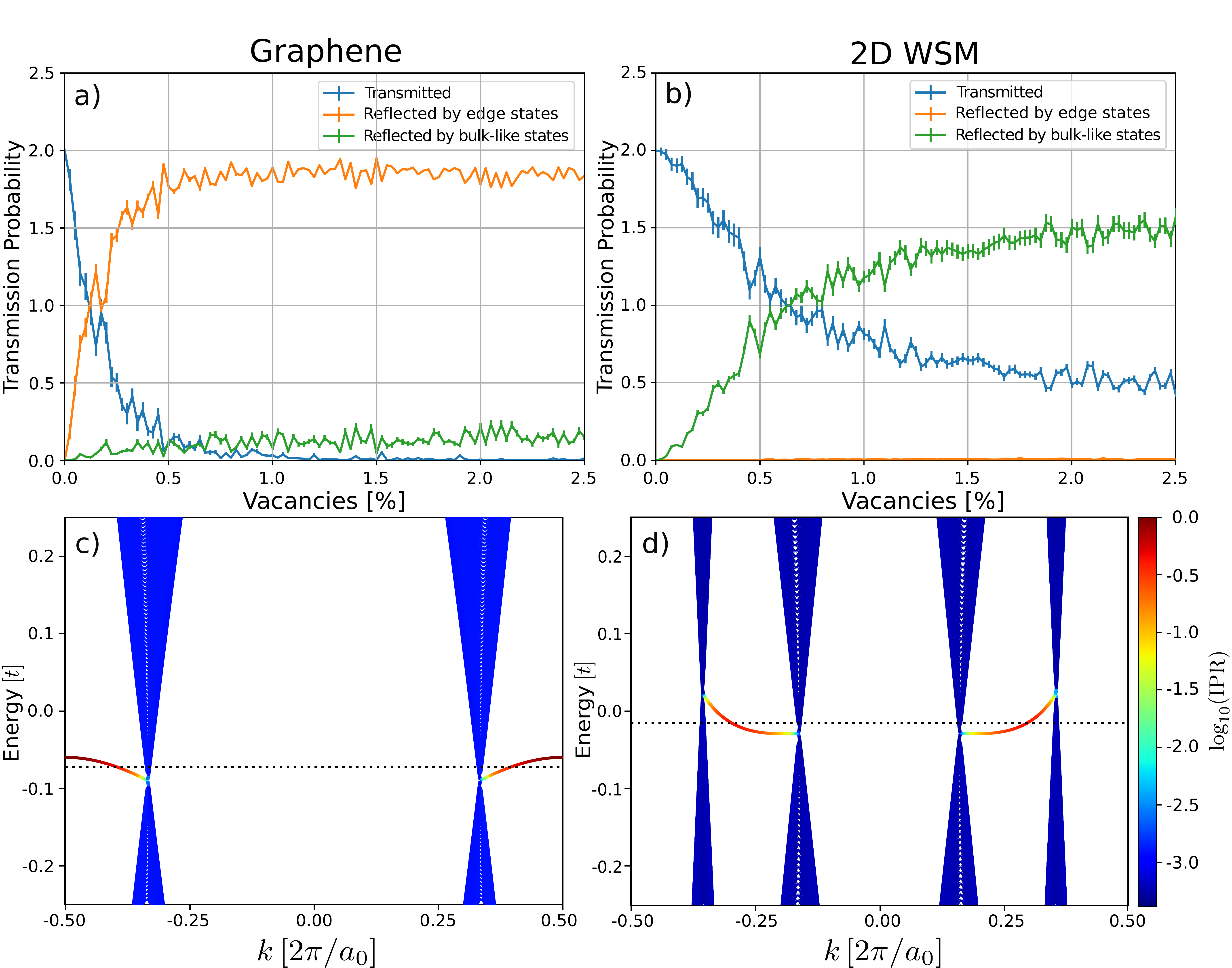}
    \caption{Transmission probability (blue) and the contributions to the reflection probability (orange for reflection through edge states and green for reflection through bulk states) for zigzag graphene and (a) and 2D WSM(b). On both panels, the probabilities are computed for incoming edge state modes (double degenerate state). The corresponding electronic band structures are shown in (c) (graphene) and (d) (2D WSM).  The parameters, $t=2$, $t_2=0.03t$ and IPR$=0.42$ is the same in both cases for a 600x400 sites system. The transmission probability showed in a) and b) correspond to the left to right propagation (according to the subset in fig.~\ref{fig:contribution}a). The reflected part has contributions from the two edge states and the bulk-like bands. The electronic band structure (c,d) shows the localization of the corresponding state encoded in the color scale. The dotted line corresponds to the energy used in the transport calculation of panels (a) and (b).}
    \label{fig:graphene}        
\end{figure*}

In general, the electronic bandstructure is the principal property when comparing both systems~\cite{matveeva2019edge,luo2020two}. Despite the edge states in graphene and 2D WSM being the similar origin (fig.~\ref{fig:edge}c), the main difference is the spin in 2D WSM and pseudo-spin in graphene. For example, the 1D Fermi arcs in 2D WSMs have a well-defined spin direction near the projection of positive monopole charge and change their sign near the opposite monopole charge projection~\cite{matveeva2019edge}. In contrast, in 2D WSMs with $\mathcal{I}$ symmetry breaking, an additional constraint related to the $\mathcal{T}$ symmetry appears. Certainly, an edge state with $k_0$ presents a well-defined spin-direction, having a $\mathcal{T}$ partner at $-k_0$ in the same edge but with opposite spin. While for graphene, the property is the same, but the pseudo-spin takes the spin role, so the behavior against impurities or disorder can be highly different. For example, against non-magnetic perturbation (\textit{i.e.} vacancies), $\mathcal{T}$ prevents the hybridization between two degenerated bands with perpendicular spin in 2D WSM, while for graphene, the mix of the pseudo-spin is allowed. Consequently, there are good reasons for believing that the 1D Fermi arcs of 2D WSM with $\mathcal{T}$  are more robust than the flat band of zigzag termination graphene, at least against perturbation that does not break $\mathcal{T}$. To prove this statement, in the next section, we analyze the transport properties of zigzag termination graphene and our 2D WSM model in the presence of random vacancies.

\section{Transport properties and vacancies effects}

To study the transport properties, we use a scattering picture for a sample of either our model 2D WSM (eq. \eqref{eq:weyl-hamiltonian}) or graphene, connected to leads (as represented by the scheme shown as inset of Fig.~\ref{fig:contribution}a) and compute the transmission coefficients using home-made Python scripts based on the Kwant package~\cite{groth2014kwant}. The contributions from different asymptotic channels allow for a more detailed interpretation of the underlying processes. For simplicity, the leads are taken identical to the sample. To assess the robustness of the different contributions to the transport properties, we use vacancies modeled through enormous onsite energy placed randomly with a uniform distribution within the sample, breaking locally inversion and all the non-symmorphic symmetries. All the transmission results are an average of twenty (or sixty for fig.~\ref{fig:contribution}) equal systems with different initial seeds in the random vacancies placement with error bars indicating the standard deviation. 

The response of bulk and edge states of 2D WSM against vacancies must first be understood before a comparison with graphene can be made, but before starting our analysis, a few remarks are in order. Without the second nearest neighbor interaction, the 1D Fermi arcs are flat bands. Furthermore, at nearby energies, the sub-bands close to the projection of Weyl nodes are abundant (the number depends on the sample's width), thereby making it difficult to distinguish the response from edge states. To help in this respect, in the simulations that follow, we consider two beneficial conditions when building the main system. First, we introduce a small dispersion to the edge states by adding a second nearest neighbor term ($t_2=0.02t$), usually occurring naturally in most materials. In addition, we consider a sample of a small width (in this case, 100 sites in $\hat{x}$-direction) to reduce the number (and influence) of bulk-like bands, showing the ability of the bulk and edge states to withstand defects. Fig.~\ref{fig:contribution}a shows the total transmission probability (blue line) and the contributions from different modes (red for modes 0 and 1, green for mode 2 and yellow for mode 3) as a function of the vacancies percent. Fig.~\ref{fig:contribution}b shows the band structure together with the numbers assigned to different modes (note that modes 0 and 1, as well as 4 and 5, are degenerate). In general,  the total transmission contains contributions from all the energy states involved in the transmission process. In our case, it can be noticed that the contributions to the total transmission curve for the pristine 2D WSM arise from only four states, two bulk-like bands, and a double degenerate edge state. Fig.~\ref{fig:contribution} also reveals exciting features as the percent of vacancies is increased. As expected, the transmission probability decays as the percentage of vacancies increases. However, it does so in a very different way for the contributions coming from bulk (modes 2 and 3) and edge states (modes 0 and 1 in red in Fig.~\ref{fig:contribution}a). Let us take, for example, the percentage of vacancies at which the transmission contribution decreases to half of its pristine value. One can see that this critical value is around $0.05\%$ for the bulk modes, while it is five times larger, about $0.25\%$, for the contribution of edge modes, revealing that edge modes are more robust to vacancies than bulk states. The origin of this difference lies in the random distribution of vacancies since it is more probably perturb a delocalized bulk state than a localized edge state with vacancies.

Now we turn to our initial trigger question: How different are the edge states in a 2D Weyl semimetal from the edge states in zigzag terminated graphene? Previous discussions have surfaced this question without providing a final answer. Here we address this by comparing our results with those obtained for zigzag graphene ribbons. To make such a comparison possible, we construct systems with similar relevant energy scales and sizes ($600\times 400$ sample sites). For graphene, we use a standard tight-binding Hamiltonian with nearest neighbors coupling $t$ and second-nearest-neighbors $t_2$.  The latter term changes the flat band in the zigzag termination for a quadratic curve (see Fig.~\ref{fig:graphene}c)) and allows for a fair comparison as the dispersion is similar to that for the 2D Weyl semimetal. The results are shown with blue lines in Fig.~\ref{fig:graphene}a for graphene and b for our 2D WSM. There is a striking difference in the robustness to vacancies. At the same time, the transmission probability decays to half its pristine value for $0.2\%$ of vacancies for graphene. In contrast, the value rises to $0.6\%$ for the 2D WSM, making it more robust than graphene against vacancies. Since both systems have identical group velocity and localization, one needs to look further to explain such a significant difference. To rationalize these results, we scrutinize the backscattering contributions. Fig.~\ref{fig:graphene}a and b show the reflection probabilities stemming either from bulk states (green) or edge states (orange) and allow us to see its differences: While for 2D WSM, edge states have a vanishing contribution to backscattering, for graphene they dominate backscattering. Bulk states contribute significantly to backscattering for 2D WSMs, but they reach lower reflection values and require much more vacancies to get to appreciable values. This striking difference is at the root of the enhanced robustness of edge states in 2D WSMs. To further explain the origin of such striking difference, one must come back to the key difference between edge states on both systems: for 2D WSMs, edge states are not spin degenerate and spin-polarized, two features absent for graphene ribbons. In other words, to achieve backscattering through edge states in 2D WSMs, the spin direction must be flipped, which is not possible through a simple vacancy defect, suppressing this backscattering channel, dominant in graphene.

\begin{figure}[t!]
    \centering
    \includegraphics[width=\columnwidth]{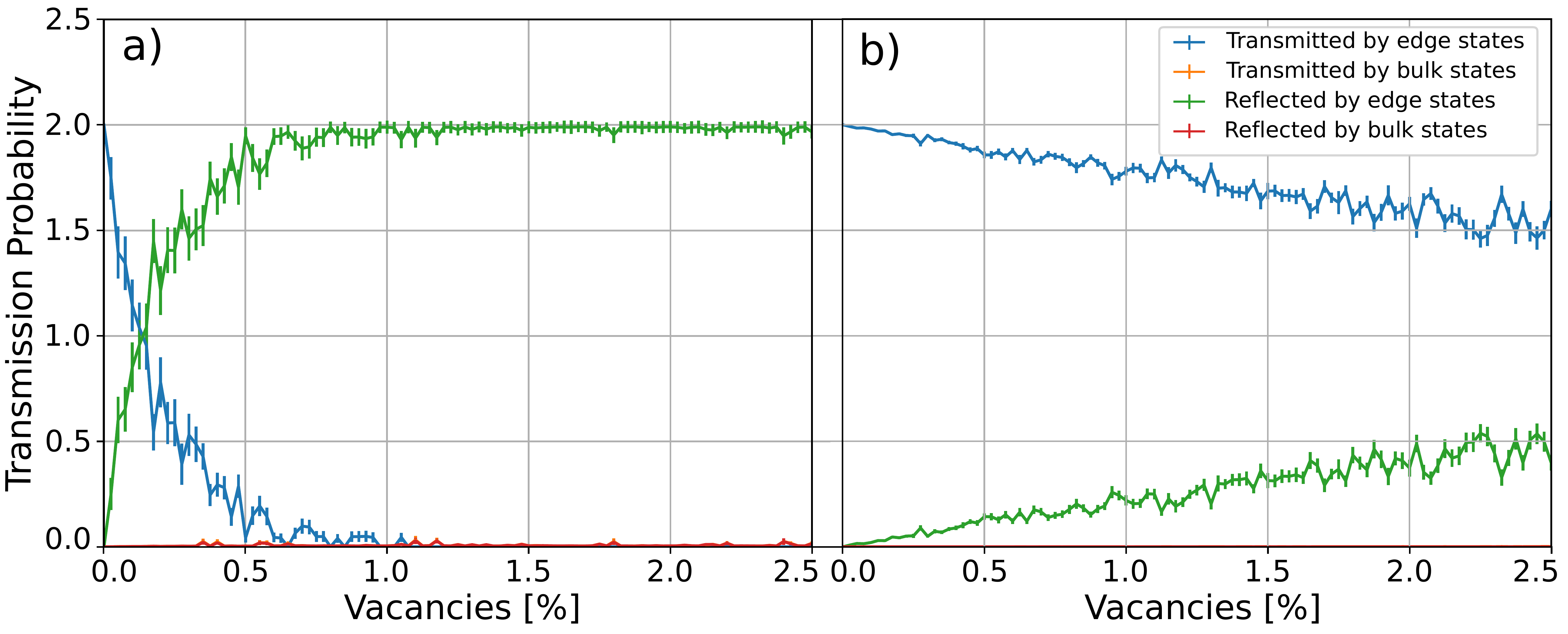}
    \caption{Transmission probability for different percentage of magnetic impurities with amplitude $m=0.05t,0.005t$ ((a) and (b) respectively) in 2D WSM. Compared with fig.~\ref{fig:graphene}b), the main difference is the presence of backscattering along the same edge due to $\mathcal{T}$ breaking term. The system has 100x1000 sites with parameters shown in fig.~\ref{fig:edge} with $t_2=0.03t$. The red and yellow line is practically zero in both panels}
    \label{fig:magnetic}        
\end{figure}

To confirm the previous argument, we present further simulations of our model 2D WSM with magnetic impurities in Fig.~\ref{fig:magnetic}. The impurities are introduced by adding a term $H_m=m\sigma_x$ in random sites using an uniform distribution, breaking $\{C_{2\hat{y}}|0\frac{1}{2}\}$ and $\mathcal{T}$ \cite{matveeva2019edge}. The transport calculation scheme used in Fig.~\ref{fig:graphene} considering an incoming wave by the edge modes is repeated for magnetic impurities. Fig.~\ref{fig:magnetic} shows that the most significant effect is that bulk-like backscattering becomes negligible, leaving backscattering through edge states as the leading mechanism. Therefore, in agreement with our previous argument, due to the $\mathcal{T}$ symmetry-breaking term, the constrain in the backscattering for bulk-like states disappears, allowing the edge backscattering. Hence, it reinforces the importance of the presence of $\mathcal{T}$ in the robustness of edge states in 2D WSM. This statement can be extended to any 2D WSM without $\mathcal{I}$, since the symmetry that does not allow backscattering between states of the same edge is $\mathcal{T}$, which is independent of the extra symmetry needed to induce a 2D WSM (or 2D DSM) phase.

One may also wonder about finite-size effects on our simulations. The main effect of reducing the system size is that vacancies can quickly build a bridge among both edges leading to backscattering, similar to what happens in other topological phases such as the quantum Hall effect. Appendix A provides results for different sample sizes.

Finally, we briefly comment on the case of 2D WSM with $\mathcal{T}$ breaking. This different case may have differences with outs from the point of view of robustness. In general, the edge states product of two Weyl nodes with different topological charges could be spin-polarized, irrespective of the $\mathcal{T}$ symmetry being preserved or broken. Matveeva \textit{et al.} shows a case of a 2D WSM with $\mathcal{T}$ symmetry breaking hosting edge states with a clear spin-polarization\cite{matveeva2019edge}. In this case, however, adding vacancies would break the ($\mathcal{I}$) symmetry required by the Weyl nodes, and thus might jeopardize the robustness of the edge states.

Summarizing, we can conclude that the edge states of 2D WSM are more robust than edge states of zigzag terminated graphene. Additionally, the presence of $\mathcal{T}$ could be the ingredient for differentiate the robustness properties of 2D WSM with $\mathcal{T}$ and $\mathcal{I}$.

\section{Final remarks}

Our study addresses two-dimensional Weyl semimetals and the robustness of the topological edge states. By using the model proposed by Young and Kane~\cite{young2015dirac} as a starting point, we add suitable symmetry-breaking terms until four Weyl nodes are isolated. The topological edge states associated with the Fermi arcs look remarkably similar to those in zigzag terminated graphene with a dispersion introduced by second nearest neighbors. Thus, it is natural to ask how similar or different is their robustness against vacancies, a question that has previously surfaced in the literature without a clear answer. Here we reveal that the edge states of two-dimensional semimetal with broken inversion symmetry behave more robustly when defects (vacancies) are introduced. The key element allowing for this robustness results from the absence of backscattering through other edge states, which in the two-dimensional Weyl semimetal have opposite spin polarization, a missing ingredient in graphene. Our results can be extended to any two-dimensional semimetal without $\mathcal{I}$ as the enhanced robustness is only due to the spin degree of freedom. The experimental perspective about this type of system could be directly related to the experimental observation of 2D DSMs with appropriate strain as to induce a 2D WSM. An example of 2D DSM that has already been synthesized is the surface-doped black phosphorus~\cite{kim2017two}, which theoretically is a 2D WSM, but the experimental results show that the splitting between the Weyl node is not big enough as to be resolved by ARPES experiments. Nevertheless, the transport properties obtained in this work could be used as a guideline for searching 2D WSM phase using the edge states, beyond the bulk band structure.

\section{Acknowledgment}

This work was funded by the National Agency for Research and Development (ANID) through grants FondeCyT \textit{postdoctorado} number 3200697 (J. D. M.) and FondeCyT Regular under grant number 1211038.  L. E. F. F. T. also acknowledges the support of The Abdus Salam International Centre for Theoretical Physics and the Simons Foundation.

\section*{References}
\bibliographystyle{apsrev4-1_title}
\bibliography{biblio}

\section*{Appendix A}

\begin{figure}[t!]
    \centering
    \includegraphics[width=1.0\linewidth]{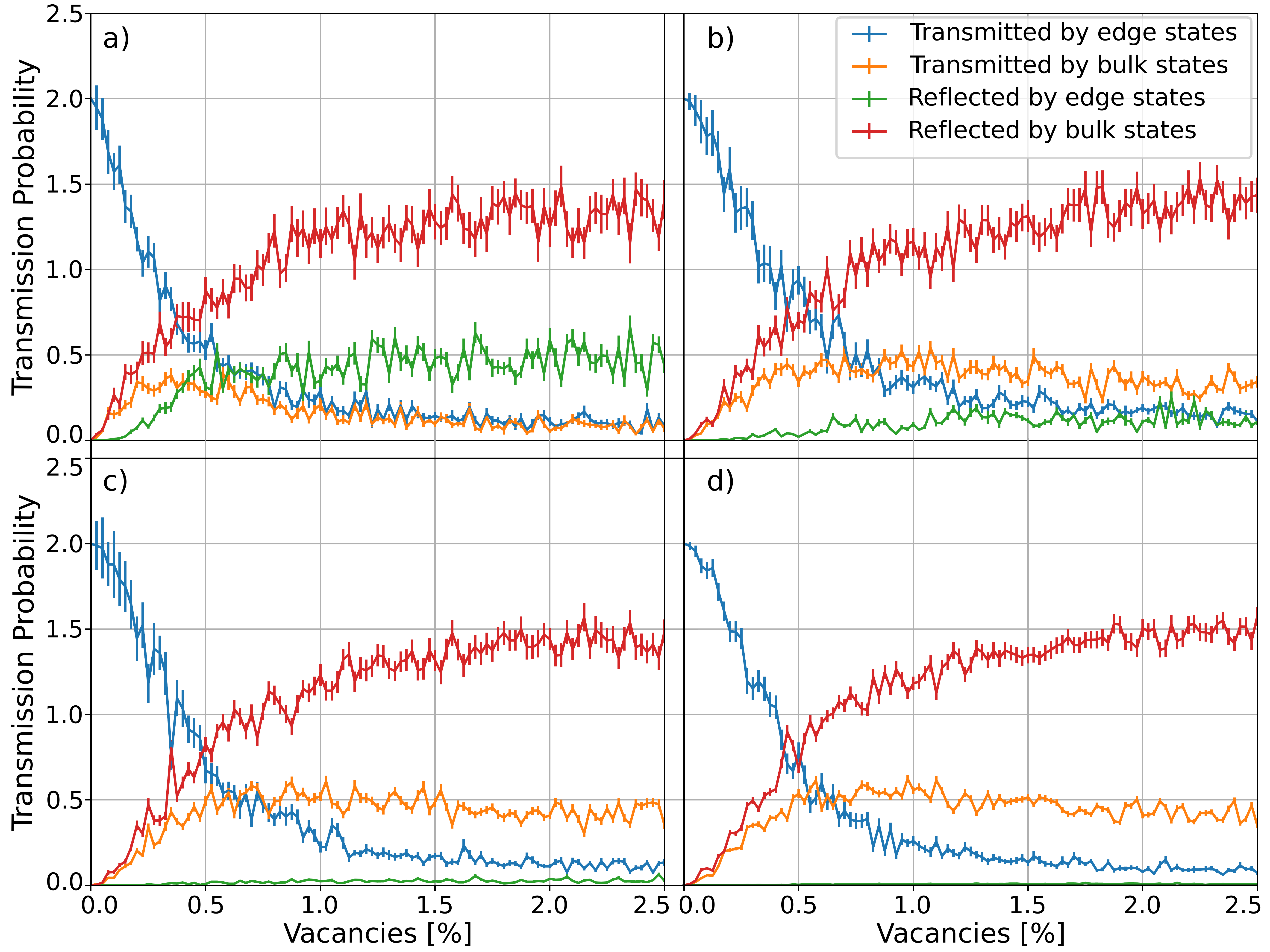}
    \caption{Transmitted and reflected part for an incident double degenerate edge states for different system sizes, a) 100x1000, b) 200x400, c) 400x400 and d) 600x400 for $t_2=0.3t$. The only difference between all the systems is the size, all the parameter and transport energy are the same.}
    \label{fig:size}        
\end{figure}

Here we provide results for different sample sizes showing that for large enough systems, the results are consistent, see Fig.~\ref{fig:size} with transport calculations for four different system sizes. For small samples, vacancies may quickly provide for a path connecting opposite edges, thereby leading to backscattering through the edge. As the system size is increased this backscattering is reduced to negligible values. In addition, the contribution to the transmission of the edge modes does not change appreciably with the system size.

\end{document}